\documentclass[doublecol]{epl2} 
\usepackage{morefloats}
\usepackage{bm}
\newcommand{\beq}{\begin{equation}}
\newcommand{\eeq}{\end{equation}}

\usepackage{comment}
\usepackage{extarrows}
\usepackage[retainorgcmds]{IEEEtrantools}
\usepackage{graphicx,placeins}
\usepackage{wasysym}
\usepackage{amsmath,amssymb,amsfonts}
\usepackage{mathtools}
\usepackage{color}
\usepackage{float}
\usepackage{txfonts}
\usepackage{nicefrac}

\usepackage{lipsum}
\usepackage[colorlinks=true,linkcolor=blue,urlcolor=blue,citecolor=blue,pdfusetitle]{hyperref}
\usepackage{physics}
\usepackage[dvipsnames]{xcolor}
\usepackage{soul}
\usepackage{ulem}
\newcommand{\mom}[1]{\langle#1\rangle}

\newcommand{\mc}[1]{\mathcal{#1}}

\newcommand{\IM}[1]{\textcolor{RoyalBlue}{#1}}

\title{Universal features of nonequilibrium Ising models in contact with two thermal reservoirs}
\shorttitle{} %Insert here a short version of the title if it exceeds 70 characters

\author{Iago N. Mamede\inst{1,2} \and Bart Cleuren\inst{2} \and Carlos. E. Fiore\inst{1}}

\institute{
\inst{1} Universidade de São Paulo, Instituto de Física, Rua do Matão, 1371, 05508-090 São Paulo, SP, Brazil\\
\inst{2} UHasselt, Faculty of Sciences, Theory Lab, Agoralaan, 3590 Diepenbeek, Belgium
}

\abstract{
We derive generic properties  of  nonequilibrium phase transitions in all-to-all Ising models placed in contact with two thermal reservoirs, in which parameters (temperatures, interactions and field parameters) assume arbitrary values depending on the contact with each thermal bath.  The presence of different kinds of external parameters leads to remarkably different sort of phase transitions. While continuous, discontinuous and even  tricritical points  are presented when external parameters are symmetric (e.g. the case of energetic barriers or different couplings between the system and thermal baths),  the tricriticality is absent when external parameters are antisymmetric  (e.g. the case of magnetic fields or biased drivings) implying that solely critical or discontinuous are possible.  In such latter case, the probability distribution acquires the Boltzmann-Gibbs like form, irrespectively the model parameters when the switching between thermal reservoirs is sufficiently  fast.  Our work sheds light about the differences between equilibrium and nonequilibrium ingredients and theirs consequences upon phase transitions.}

\begin{document}
\maketitle

\section{Introduction}
Nonequilibrium phase transitions are ubiquitous in nature and can exhibit remarkable differences in relation with their equilibrium counterparts. They manifest in several systems
in the scope of classical  \cite{odor2008universality}, quantum systems \cite{PhysRevLett.124.170603}, chemistry \cite{marro2005,fiore2021current,basile}, biology ~\cite{rapoport1970sodium, gnesotto2018broken,lynn2021broken,smith2019public}, sociophysics \cite{castellano2009statistical} and others. In a similar
fashion to equilibrium counterparts, they are often characterized via an order parameter, but
unlike them, they are featured by persistent probability currents stemming from different drivings forces, such as
different thermal baths~\cite{mamede2025collectiveheatenginesdifferent}, ``biased" external control parameters (e.g. energetic barriers and drivings forces )~\cite{mamede2023,forao2023powerful,gatien,mamede2023} or time dependent parameters \cite{mamede2021obtaining,noa2020thermodynamics,rosas1}.

Since there is no toolbox like equilibrium thermodynamics
for characterizing nonequilibrium systems, different approaches have been proposed and adopted \cite{seifert2012stochastic,crooks,barato2015thermodynamic,karel2016prl}. In particular, recent attention has been devoted to the role of the entropy production—a key indicator of system irreversibility— in order to understand nonequilibrium phase transitions \cite{PhysRevLett.108.020601, noa2019, aguilera2023nonequilibrium, martynec2020entropy}. Despite this, several issues still remain open. Notably, entropy production behavior is often characterized by non-universal features, and its relationship with standard critical exponents is relatively unknown \cite{martynec2020entropy, mamede2025collectiveheatenginesdifferent,noa2019entropy}. Very recently,  a closed form for the probability distribution of nonequilibrium systems simultaneously placed in contact with two thermal reservoirs has been obtained\cite{mamede2025exactmappingnonequilibriumequilibrium}, providing an alternative recipe for characterizing nonequilibrium phase transitions, via the connection with the equilibrium  case.

In the last decades,  the Ising model has served as a paradigmatic and ideal laboratory for addressing different properties of phase transitions and collective phenomena. Although introduced and extensively studied in the scope of equilibrium statistical mechanics, its simplicity spans extensions beyond the equilibrium,
ranging from phase transitions 
\cite{Garrido1987,Garrido1989, Blote1990,Tome_1991,PhysRevLett.108.020601, Gambetta2019,aguilera2023nonequilibrium,Yan2023,dutta2025},
biology \cite{pnas21} and
voter models
\cite{harunari2017partial, encinas2018fundamental,encinas2019majority, fiore23,felipe1,fiore25} to social
dynamics \cite{castellano2009statistical} and heat engines \cite{gatien,mamede2023}. 

{Motivated by such queries, we investigate the phase transition properties 
and critical behavior f nonequilibrium Ising models in contact with two different thermal reservoirs.  We propose two descriptions for the thermal bath contact and the influence of external parameters: simultaneous and non simultaneous contact between reservoirs as well as  
the influence of antisymmetric and symmetric external parameters. While the former describe  magnetic fields ~\cite{mamede2025exactmappingnonequilibriumequilibrium,PhysRevE.87.052123,PhysRevE.59.2710} or  biased forces favoring certain transitions \cite{gatien},  symmetric parameters appear in  transitions that depend on energetic barrier between states ~\cite{RevModPhys.62.251,Hovorka_2017,10.1103/physrevb.89.104410,C7RA07734C,10.1063/1.1772358}.
%The first class corresponds to antisymmetric field terms, which may play roles analogous to external magnetic fields or to biased forces that depend simultaneously on transitions and bath properties~\cite{gatien,PhysRevResearch.5.043278,hzl3-hjnl,PhysRevLett.109.190602,Schmiedl_2008}. The second class consists of symmetric, transition-independent terms, naturally interpreted as barrier-like contributions in chemical-process models or, alternatively, as ingredients that in lattice systems may lead to tricritical behavior in the same way to that found in metamagnets~\cite{10.1103/physrevb.84.134422,10.1016/j.jmmm.2007.10.006}.
Our analysis is carried out for all-to-all descriptions~\cite{gatien, mamede2023} in which the system is stochastically coupled to alternating hot and cold thermal baths via a finite and symmetric switching rate \cite{mamede2025exactmappingnonequilibriumequilibrium,
mamede2025collectiveheatenginesdifferent,busiello2020coarse,forao2025universal}, whose regime of fast switchings recovers the simultaneous contact between thermal baths.  }
%The paper is organized as follows. In Sec.~\ref{Model}, we introduce the model, beginning with its microscopic formulation and subsequently describing its mesoscopic dynamics. The thermodynamic limit and the corresponding large-deviation treatment are presented in Sec.~\ref{Large_N}. In Sec.~\ref{Fast_Swi_Sec}, we analyze the regime of fast switching between the thermal baths, analyzing separately the different types of external control terms and their associated phase-transition features. The case of finite switching rates is examined in Sec.~\ref{Finite_Hoop}. We conclude with a summary and outlook in Sec.~\ref{Conclusion}.

The different kinds of external parameters give rise to remarkably distinct phase transitions. While they can be critical, discontinuous, and tricritical points for symmetric parameters,  the tricriticality is absent for antisymmetric ones. In such latter case, the limit
of fast-switchings lead to a Boltzmann–Gibbs–like distribution regardless of model details.

\section{Model and all-to-all dynamics}
\begin{figure}[htb!]
 \includegraphics[width=.95\columnwidth]{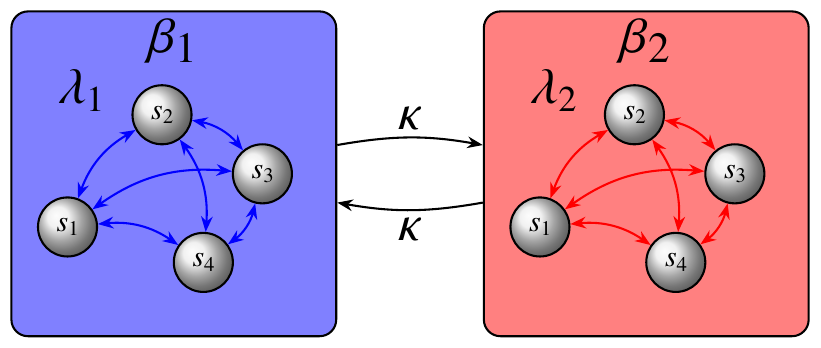} 
 \caption{Schematics of a model composed of 4 spins with an arbitrary connectivity. The system is connected to a hot or a cold thermal bath, with reciprocal temperatures $\beta_2$ and $\beta_1$, respectively, that stochastically exchange with each other with rate $\kappa$.}
 %\caption{Schematics of model: The dynamics are composed of one-site transitions,
 %from   $\bm{s}^{(\nu)}_i = \{s^{(\nu)}_1,s^{(\nu)}_2,...,s^{(\nu)}_i\ldots,s^{(\nu)}_N\}$  to $\bm{s}^{(\nu)}_j = \{s^{(\nu)}_1,s^{(\nu)}_2,...,-s^{(\nu)}_i\ldots,s^{(\nu)}_N\}$   and  stochastic hoppings with $\kappa$  between thermal baths at the same $\bm{s}^{(\nu)}_i$. Parameter $\beta_\nu$ and $\lambda_\nu$ denote the reciprocal temperature and the strength of  external  parameter at each subsystem, respectively.}
 \label{scheme}
 \end{figure}

 The Ising model is described by a collection of $N$ interacting units with connectivity $k$, each one can  be in a state $ s_i\in\{-1,+1\}$ ($\bm{s} = \{s_1,\ldots,s_N\}$) due to the contact with the $\nu$-th bath, with $\nu\in\{1,2\}$. Hence, each configuration  is described by the set $(\bm{s},\nu)$,  and at any given time, the system is entirely coupled to a single reservoir, and the switching process globally alternates the thermal environment. The energy for the system is given by
\begin{equation}
\mc{E}(\bm{s})=-\frac{\epsilon}{2k}\sum_{i,j\neq i}s_is_{j},
\label{General_Energy}
\end{equation}
where $\epsilon$ denotes the system configuration and interaction strength between two spins, respectively. The dynamics is characterized by two kind of transitions: (i) spin flips, while coupled to the same reservoir  at the  (reciprocal) temperature $\beta_\nu=(k_BT_\nu)^{-1}$, with rate $W^{(\nu)}({\bm{s}\to\bm{s}'})$, from the configuration $(\bm{s},\nu)$ to the $(\bm{s}',\nu)$, or (ii) reservoir changes with fixed spin configuration with rate $\kappa$, from $(\bm{s},\nu)$ to $(\bm{s},\nu')$ ($\nu\neq\nu'{\rm and}\in\{1,2\}$). Additionally, the influence of an external parameter, namely $\lambda_1\text{ and }\lambda_2$, associated to each thermal bath, will be included into the dynamics.
%\CEF{For
%simplicity,}
%characterized by its orientation ($\pm$) and by the bath to
%which it is currently coupled $(\nu=1,2)$.
The schematics of model and different parameters are shown in Fig.\ref{scheme}.

%\begin{align}
%    \mc{J}_{ij}
%    &=W(\bm{s}_i,\bm{s}_j)P(\bm{s}_j,t)-W(\bm{s}_j,\bm{s}_i)P(\bm{s}_i,t),
%\end{align}

%s mentioned, the transition mechanisms split into two classes. The first corresponds to spin flips within a fixed reservoir:
 %The second class corresponds to bath-switching processes, where the spin configuration is preserved just switching to the correspondent set of states in the other subset such that $ W(\bm{s}^{(\nu)}_i,\bm{s}^{(\nu)}_j)=\kappa$ for $\bm{s}_j=\bm{s}^{(1)}_i,\bm{s}_i=\bm{s}^{(2)}_i$ or the other way around. These exchanges are assumed to be independent of the spin state, depending only on the initial subset of $\mc{S}$.

In order to grasp our main goals (comparison between simultaneous versus non simultaneous thermal bath contact and symmetric versus antisymmetric parameters) we focus our analysis  on
fully connected networks, commonly referred as  all-to-all interactions, where each unit interacts with all others.  In such case,  by setting $k\rightarrow N$, each configuration gets fully characterized by the number of sites in each state  here given by the vector $\bm{n}=(n_{-},n_{+})$~\cite{RevModPhys.97.015002}, with the constraint $n_{-}+n_{+}=N$, along with the refereed bath $\nu$.  The system energy
 $\mc{E}(\bm{n})$ is then given by
\begin{equation}
    \mc{E}(\bm{n})=-\frac{\epsilon}{2N}\left[\sum_{\alpha =\pm} n_{\alpha}(n_{\alpha}-1) - 2n_{-}n_{+} \right].
    \label{eq2}
\end{equation}

\begin{comment}
\IM{We formulate the stochastic dynamics in terms of
an effective chemical reaction network. For that, we distinguish spins according
to the thermal bath to which they are currently coupled and introduce
four effective species $X_-^{(1)}, X_+^{(1)},
X_-^{(2)},X_+^{(2)},$
representing spins in states $-1$ and $+1$ interacting with bath
$\nu=1,2$, respectively, where the aforementioned $n_\alpha^{(\nu)}$ is the population number of species
$X_\alpha^{(\nu)}$ with $\alpha\in\{-,+\}$.}\IM{Within this description, the stochastic dynamics consists of
elementary reactions corresponding to spin flips within each bath
and stochastic hopping between baths while preserving the spin state. The possible reactions in this description yields}
\end{comment}

{
The probability distribution $P(\bm{n},\nu)$ then evolves according
to the master equation~\cite{gardiner2010stochastic}
\begin{align}
\label{Meso_ME}
\partial_t P(\bm{n},\nu)
&=
\sum_{r\in\{-,+\}} W^{(\nu)}_r(\bm{n}-\bm{\Delta}_r)P(\bm{n}-\bm{\Delta}_r,\nu)
-
W^{(\nu)}_r(\bm{n})P(\bm{n},\nu)\\
&+\kappa[P(\bm{n},\nu')-P(\bm{n},\nu)]\nonumber
\end{align}
where for transitions between different
states at the same temperature $r\in\{-,+\}$ labels the elementary transitions and
$\bm{\Delta}_r$ denotes the corresponding transition vectors with $\bm{\Delta}_{+}=(-1,+1)$ and $\bm{\Delta}_{-}=-\bm{\Delta}_{+}$,
whereas the second term in Eq.\eqref{Meso_ME} accounts to transitions
between different thermal reservoirs $\nu'\neq\nu$.
Following the usual mass–action form of
chemical reaction networks~\cite{VANDERSCHAFT201324}, the rates $W^{(\nu)}_r(\bm{n})$ are expressed by $W^{(\nu)}_{+}(\bm{n}) = \omega^{(\nu)}_{\bm{n}+\bm{\Delta}_{+},\bm{n}}\, n_{-}$ and $W^{(\nu)}_{-}(\bm{n}) = \omega^{(\nu)}_{\bm{n}+\bm{\Delta}_{-},\bm{n}}\, n_{+}$.

%\IM{\begin{align}
%W_{1,2}(\bm{n}) = %\omega^{(\nu)}_{\bm{n}+\bm{\Delta}_{1,2},\bm{n}}\, n_{\mp}, 
%\end{align}}
Throughout this work, we shall define transition
rates between states according to Arrhenius form 
\begin{equation}
\omega^{(\nu)}_{\bm{n}+\bm{\Delta}_r,\bm{n}}=\gamma e^{-\frac{\beta_\nu}{2}\left[\mc{E}(\bm{n}+\bm{\Delta}_r)-\mc{E}(\bm{n})+d^{(\nu)}(\bm{\Delta}_r)\lambda_\nu\right]},
\end{equation}
where accounts to the contribution
of an external parameter (e.g. magnetic field, non-conservative driving or a energetic barrier) with signal
$d^{(\nu)}(\bm{\Delta}_r)\in\{-1,1\}$}. Two kinds of external parameters
will be investigated, in which
$d^{(\nu)}(\bm{\Delta}_r)$ is symmetric 
$d^{(\nu)}(\bm{\Delta}_{r})=d^{(\nu)}
(-\bm{\Delta}_{r})$ 
or antisymmetric
$d^{(\nu)}(\bm{\Delta}_r)=-d^{(\nu)}
(-\bm{\Delta}_{r})$. 

\section{Large $N$ limit and phase transitions}\label{Large_N} 
As mentioned previously, the system dynamics occurs through two different mechanisms: the spin flip and global exchange of thermal baths.  The limit
$N\rightarrow \infty$ is of particular interest not only because
quantities become simpler but also it gives rise to phase transitions
and collective behavior.
%Hence, both processes can be treated as two independent processes, with the joint probability distribution being decomposed as $P(\bm{n},\nu)=P(\bm{n}|\nu)P(\nu)$, with $P(\bm{n}|\nu)$ the conditioned probability distribution (CPD) and $P(\nu)$ the one correspondent for the system be in contact with bath $\nu$.}
%\IM{By using $\partial_tP(\bm{n}|\nu)=[\partial_tP(\bm{n},\nu)-P(\bm{n}|\nu)\partial_tP(\nu)]/P(\nu)$ along with the expression yielded by the marginalization condition $\partial_tP(\nu)=\kappa[P(\nu')-P(\nu)]$ and the symmetry of baths $P(\nu)=P(\nu')$, we recover the master equation for the CPD as an analogous format with Eq.\eqref{Meso_ME} given by }
%\IM{\begin{align}
%\partial_t P(\bm{n}|\nu)
%&=
%\sum_r J_r(\bm{n}|\nu)+\kappa[P(\bm{n}|\nu')-P(\bm{n}|\nu)],
%\label{Meso_ME}
%\end{align}}
%\IM{with $J_r(\bm{n}|\nu)=W_r(\bm{n}-\bm{\Delta}_r)P(\bm{n}-%\bm{\Delta}_r|\nu)
%-
%W_r(\bm{n})P(\bm{n}|\nu)$, the conditioned probability current.}
%Our interest lies in the  limit of $N\to\infty$ in which different  phase transitions take place. 
For that, it is convenient to introduce the density of state
$\pm$ associated to each thermal reservoir,
$x^{(\nu)}_{\pm}=\langle n_{\pm}^{(\nu)}\rangle/N$,
%$x^{(\nu)}_{\pm}:=\sum_{\bm{n}}n_{\pm}P(\bm{n}|\nu)/N$, 
in such a way the dynamics is characterized by the vector
  $\bm{x}\equiv(x^{(1)}_{-},x^{(1)}_{+},x^{(2)}_{-},x^{(2)}_{+})$. 
%\begin{equation}
%\langle \dot n_i \rangle = \sum_{j\neq i} \big[\langle J(n_i,n_j)\rangle - \langle J(n_j,n_i)\rangle\big].
%\tag{19}
%\end{equation}
Since we are dealing with a MFT like description
as $N\rightarrow \infty$, one can resort to the ideas
of Refs. \cite{VANKAMPEN2007193,herpich} in which
each average of type 
$\langle \omega^{(\nu)}_{\bm{n}+\bm{\Delta}_{+},\bm{n}}\, n_{\pm}\rangle$ can be replaced as
$\langle \omega^{(\nu)}_{\bm{n}\pm \bm{\Delta}_{+},\bm{n}}\, n_{\pm}\rangle\approx \langle\omega^{(\nu)}_{\bm{n}\pm\bm{\Delta}_{+},\bm{n}}\rangle \langle n^{(\nu)}_{\pm}\rangle\rightarrow \omega^{(\nu)}_{+-} x^{(\nu)}_{\pm}$
where
\begin{align}
    \omega^{(\nu)}_{+-}&=\gamma \exp\left\{-\frac{\beta_\nu}{2}\left[-2\epsilon( x_{+}^{(\nu)}-x_{-}^{(\nu)})+d^{(\nu)}_{+-}\lambda_\nu\right]\right\},\label{Omega_pm}\\
    \omega^{(\nu)}_{-+}&=\gamma \exp\left\{-\frac{\beta_\nu}{2}\left[2\epsilon  ( x_{+}^{(\nu)}-x_{-}^{(\nu)})+d^{(\nu)}_{-+}\lambda_\nu\right]\right\}\label{Omega_mp}.
\end{align}
%and the system energy is given by
%\begin{equation}
%    \widetilde{\mc{E}}^{(\nu)}(\bm{x})=-\frac{\epsilon}{2}\left(x_{+}^{(\nu)}-x_{-}^{(\nu)}\right)^2\label{macro_Energy}.
%\end{equation}
The time evolution of the each mean density of occupation reads
\begin{equation}
 \begin{cases}
        \partial_tx^{(1)}_{-}=\omega_{-+}^{(1)}x^{(1)}_{+}-\omega_{+-}^{(1)}x^{(1)}_{-}+\kappa (x^{(2)}_{-}-x^{(1)}_{-}),\\
        \partial_tx^{(1)}_{+}=\omega_{+-}^{(1)}x^{(1)}_{-}-\omega_{-+}^{(1)}x^{(1)}_{+}+\kappa (x^{(2)}_{+}- x^{(1)}_{+}),\\
        \partial_tx^{(2)}_{-}=\omega_{-+}^{(2)}x^{(2)}_{+}-\omega_{+-}^{(2)}x^{(2)}_{-}+\kappa (x^{(1)}_{-}-x^{(2)}_{-}),\\
       \partial_tx^{(2)}_{+}=\omega_{+-}^{(2)}x^{(2)}_{-}-\omega_{-+}^{(2)}x^{(2)}_{+}+\kappa( x^{(1)}_{+}-x^{(2)}_{+}).
\end{cases}
\end{equation}
By defining the order parameter for each subsystem $\nu$ as $m_\nu(t)=x^{(\nu)}_{+}(t)-x^{(\nu)}_{-}(t)$ and the global magnetization $m(t)=m_1(t)+m_2(t)$,
the steady-state regime, in which $m_\nu(t)\to \overline{m}_\nu$ and $m(t)\to\overline{m}$, is given by the set of  non-linear coupled equations 
    \begin{align}
       \overline{m}_1&=\frac{\sinh \left(H_1\right)
   \left[\kappa  e^{\beta _2 \delta _2
   \lambda _2}+2 \gamma  \cosh
   \left(H_2\right)\right]+\kappa 
   \sinh \left(H_2\right) e^{\beta _1
   \delta _1 \lambda _1}}{\cosh
   \left(H_1\right) \left[\kappa 
   e^{\beta _2 \delta _2 \lambda _2}+2
   \gamma  \cosh
   \left(H_2\right)\right]+\kappa 
   \cosh \left(H_2\right) e^{\beta _1
   \delta _1 \lambda _1}}\label{sol1_Mag}\\
   \overline{m}_2&=\frac{\sinh \left(H_2\right)
   \left[\kappa  e^{\beta _1 \delta _1
   \lambda _1}+2 \gamma  \cosh
   \left(H_1\right)\right]+\kappa 
   \sinh \left(H_1\right) e^{\beta _2
   \delta _2 \lambda _2}}{\cosh
   \left(H_1\right) \left[\kappa 
   e^{\beta _2 \delta _2 \lambda _2}+2
   \gamma  \cosh
   \left(H_2\right)\right]+\kappa 
   \cosh \left(H_2\right) e^{\beta _1
   \delta _1 \lambda _1}}\label{sol2_Mag}
    \end{align}
where $H_\nu\equiv\beta_\nu(\epsilon m_\nu-\theta_\nu\lambda_\nu)$ and  parameters
$\delta_\nu\equiv[d^{(\nu)}_{+-}+d^{(\nu)}_{-+}]/4$, $\theta_\nu\equiv[d^{(\nu)}_{+-}-d^{(\nu)}_{-+}]/4$ have been conveniently introduced
in order to analyze two opposite set of external parameters, $\delta_\nu=0$  and $\theta_\nu=0$, the former and latter from now on referred as antisymmetric and symmetric cases, respectively.  Eqs.~(\ref{sol1_Mag}) and (\ref{sol2_Mag}) are valid for any kind
of external parameters $\lambda_\nu$'s, temperatures $\beta_\nu$'s
and switching rates $\kappa$'s. Also,
the steady entropy production, a key indicator of the system irreversibility \cite{noa2019entropy,martynec2020entropy,fiore2021current,goes20}, acquires (in all cases) a simple form, derived in Appendix, and given by 
\begin{equation}
    \mom{\dot{\sigma}}=\sum_\nu \overline{H}_\nu(\omega^{(\nu)}_{+-}-\omega^{(\nu)}_{-+})(1-\overline{m}_\nu\coth\overline{H}_\nu),
    \label{EP_macro}
\end{equation}
where  $\overline{m}_\nu$ is given by the solution of Eqs.~\eqref{sol1_Mag}-\eqref{sol2_Mag}.

\section{Fast switchings between thermal baths}\label{Fast_Swi_Sec}
 The limit of rapid exchange between the baths, i.e., for $\kappa\to\infty$,
presents remarkable features, not only because quantities become simpler,
 but also they have been studied in different previous systems \cite{cleuren2001ising,gatien,mamede2023,forao2023powerful,forao2024splitting}.
  In such case, the dynamics of the two coupled order parameters 
  acquires the single form $\overline{m}_\nu \to \overline{m}/2$.\begin{comment}
The steady state probability distribution $\pi(\overline{m})$ is evaluated through
%\begin{equation}
%    \pi(\overline{m})=\frac{1}{Z}\exp\left[-N\int_{m_0}^{\overline{m}}d\overline{m}'~p(\overline{m}')\right].
        $\pi(\overline{m})\sim \exp\left[-NS(\overline{m})\right]$,
%    \label{Steady_State_P_LD}
%\end{equation} where
where $S(\overline{m})=\int_{m_0}^{\overline{m}}d\overline{m}'~p(\overline{m}')$ and  $p(\overline{m})$  obtained from Eq.~\eqref{HJ_equation} as $\mc{H}=0$. Despite long (its full expression  shown in Appendix), it is     
valid for a general kind of external parameter, $\delta_\nu$ and $\theta_\nu$. 
\end{comment}
The entropy production $ \mom{\dot{\sigma}}$ acquires a simpler form, given by
\begin{equation}
\label{epp}
    \mom{\dot{\sigma}}=\frac{2 \gamma  (\overline{H}_1-\overline{H}_2) \sinh
   \left(\overline{H}_1-\overline{H}_2\right)}{e^{\beta _2 \delta _2 \lambda _2}\cosh \overline{H}_1
   +e^{\beta _1 \delta _1 \lambda _1}\cosh  \overline{H}_2
   },
\end{equation}
and vanishes as $\overline{H}_1=\overline{H}_2$, consistent with $\beta_1=\beta_2$, $\lambda_1=\lambda_2$ and $\theta_1=\theta_2$ being
$>0$ otherwise, provided ${\overline m}\neq 0$.

\subsection{Antisymmetric external parameters}\label{asy}
\begin{figure}
    \centering
    \includegraphics[width=.95\linewidth]{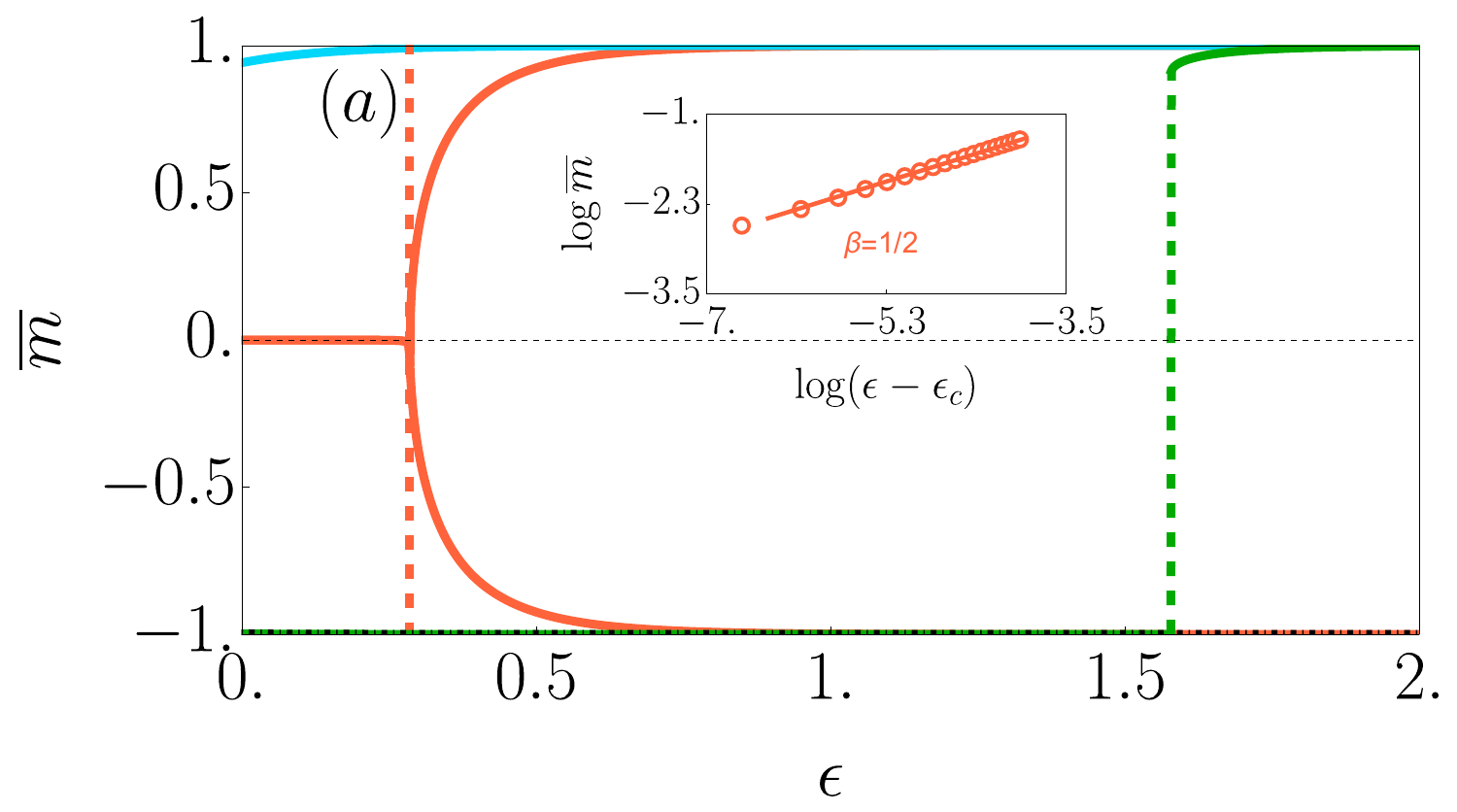}
    \includegraphics[width=.95\linewidth]{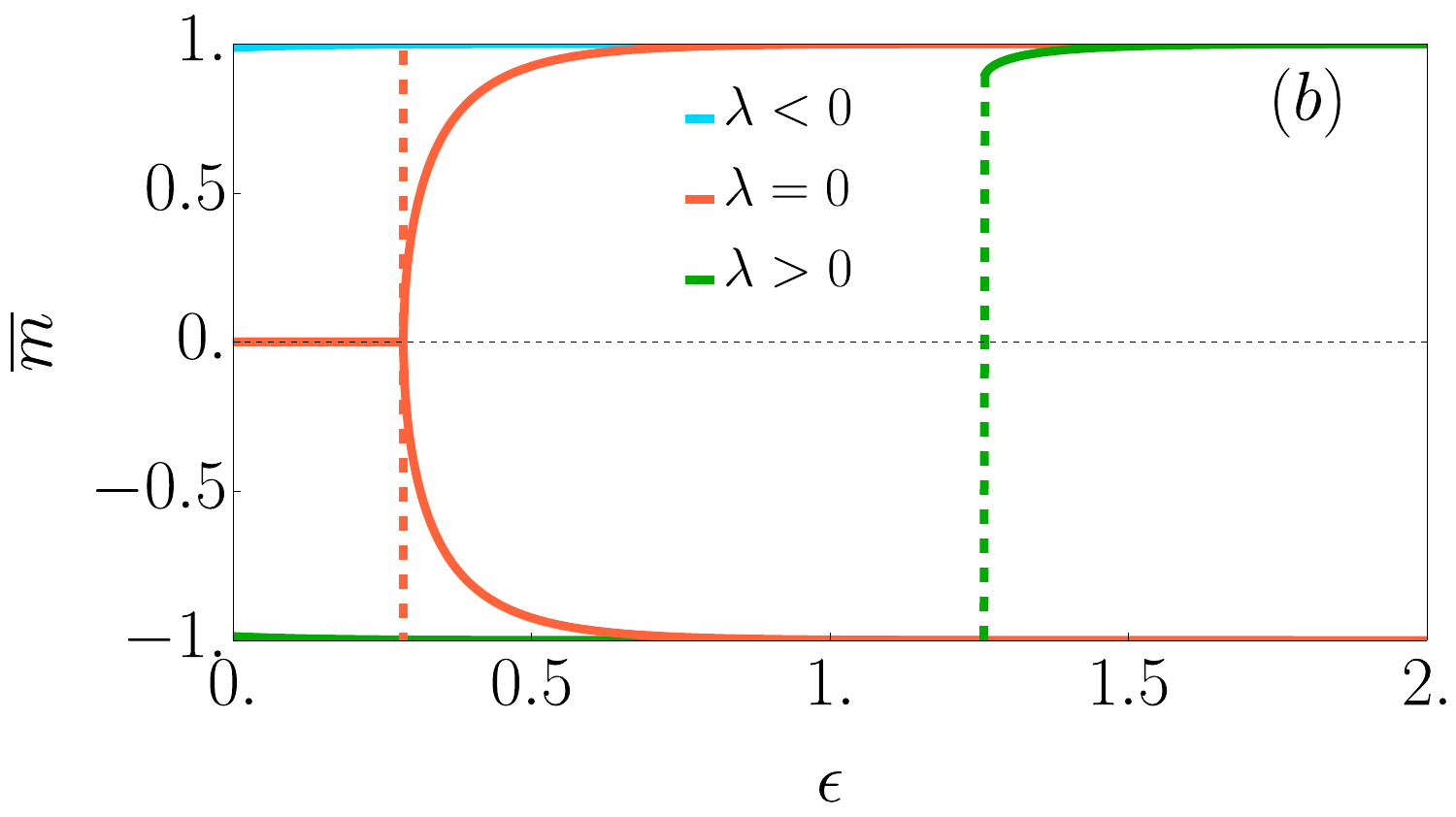}
    \caption{The behavior of Ising model for
    antisymmetric external parameters ($\delta_1=\delta_2=0$) and different $\lambda$'s ($\lambda_1=\lambda_2=\lambda=0,\pm1$). Top and  bottom panels
    depict the total magnetization ${\overline m}$ for
    $\theta_1=\theta_2=1$ and $\theta_1=-\theta_2=1$, respectively. Vertical
    dashed lines denote the corresponding transition points. Inset:
   Log-log plot of $\overline{m}$ versus $\epsilon-\epsilon_c$, consistent with scaling $\overline{m}\sim(\epsilon-\epsilon_c)^{1/2}$. Parameters: $\beta_1=6,\beta_2=1\text{ and }\gamma=1$}
    \label{mag_Anti}
\end{figure}

For  fully antisymmetric
external parameters, $\delta_\nu=0$ and fast switchings,
the total magnetization ${\overline{m}}={\overline{m}}_1+{\overline{m}}_2$ acquires a Curie-Weiss
like equation
\begin{equation}
    \overline{m}
    =\tanh\!\left[
    \frac{(\beta_1+\beta_2)\epsilon}{2}\,\overline{m}
    -\frac{\beta_1\theta_1\lambda_1+\beta_2\theta_2\lambda_2}{2}
    \right].
    \label{cw}
\end{equation}
We firstly note that it reduces
to the well-known Curie--Weiss mean-field relation when $\theta_1\lambda_1=\theta_2\lambda_2=-h$ and $\beta_1=\beta_2$ ~\cite{salinas2001introduction}. From this connection, it is immediate to see 
the  coexistence between two ferromagnetic phases
takes place when $(\beta_1+\beta_2)\epsilon>2$ and $  \beta_1\theta_1\lambda_1+\beta_2\theta_2\lambda_2=0$, whereas
no phase transition as $(\beta_1+\beta_2)\epsilon<2$. The criticality then yields at
\begin{equation}
    \begin{cases}
        (\beta_1+\beta_2)\,\epsilon_c = 2,\\[2mm]
        \beta_1\theta_1\lambda_1+\beta_2\theta_2\lambda_2 = 0,
    \end{cases}
    \label{cond_Anti}
\end{equation}
irrespectively the values of $\theta_\nu$ and $\lambda_\nu$. In order to obtain the
critical exponents, we  expand Eq.~(\ref{cw}) into
power series given by $0=A_0+A_1(\epsilon-\epsilon_c)\overline{m}+A_2\overline{m}^2+A_3\overline{m}^3+\dots$, whose coefficients
$A_i$'s are shown in the Appendix.
From the criticality condition given by Eq.~(\ref{cond_Anti}), it follows that $A_0=A_2=0$ and $A_3<0$ and  the order parameter behaves as 
$\overline{m}\sim(\epsilon-\epsilon_c)^{\beta_c}$, where $\beta_c=1/2$, akin to the behavior of Ref.~\cite{mamede2025exactmappingnonequilibriumequilibrium}. 
  A remarkable  feature about the antisymmetric external parameters comes
  from the fact that the probability  distribution
assumes Boltzmann-Gibbs like form, irrespective the temperatures and external parameters.
To see this,  we take the case of finite $N$, in which the
probability distribution is given by
\begin{equation}
\label{steady_state_Anti}
    \pi(\bm{n})=\left(\frac{N!}{n_{+}!n_{-}!}\right)\exp\left\{-\frac{1}{2}\sum_\nu\left[\beta_\nu(\mathcal{E}(\bm{n})+\theta_\nu\lambda_\nu(n_{+}-n_{-}))\right]\right\}.
\end{equation}
By rewriting the combinatorial term as $N\log N-\sum_{r=\pm}n_r\log n_r$ as $N,n_r\gg1$ and taking $N\rightarrow \infty$, one gets the
previous expression for $\pi({\overline m})\propto \exp[-N\mathcal{S}(\overline{m})]$, where
$ \mathcal{S}(\overline{m})$ is given by
   \begin{eqnarray}
   \label{s}
        \mathcal{S}(\overline{m})&=&-\frac{\epsilon(\beta_1+\beta_2)}{4}\overline{m}^2+\frac{(\beta_1\theta_1\lambda_1+\beta_2\theta_2\lambda_2)}{2}\overline{m}\nonumber\\&-&\sum_{r=\pm}\left(\frac{1+r\overline{m}}{2}\right)\log\left(\frac{1+r\overline{m}}{2}\right).
    \end{eqnarray}
By maximizing Eq.~(\ref{s}) with respect to $\overline{m}$ as $N\rightarrow \infty$, one recovers Eq.~(\ref{cw}).
   It is worth mentioning that Eq.~(\ref{s}) extends the findings from \cite{mamede2025exactmappingnonequilibriumequilibrium} to 
the case of $\lambda_\nu$ playing the role of a generic parameter.
We close this section by exemplifying in Fig. \ref{mag_Anti} the order-parameter versus $\epsilon$  for
a continuous and discontinuous phase transitions for different kinds
of antisymmetric external parameters,
$\theta_1=\theta_2$ (top) and  $\theta_1=-\theta_2$ (bottom),
depicting critical behaviors consistent with $\beta_c=2$ as Eq.~ (\ref{cond_Anti}) is fulfilled as well as the existence
of discontinuous phase transitions as $(\beta_1+\beta_2)\epsilon>2$,
irrespective the form of external parameter.

\begin{figure}[htb!]
    \centering
    \includegraphics[width=.95\linewidth]{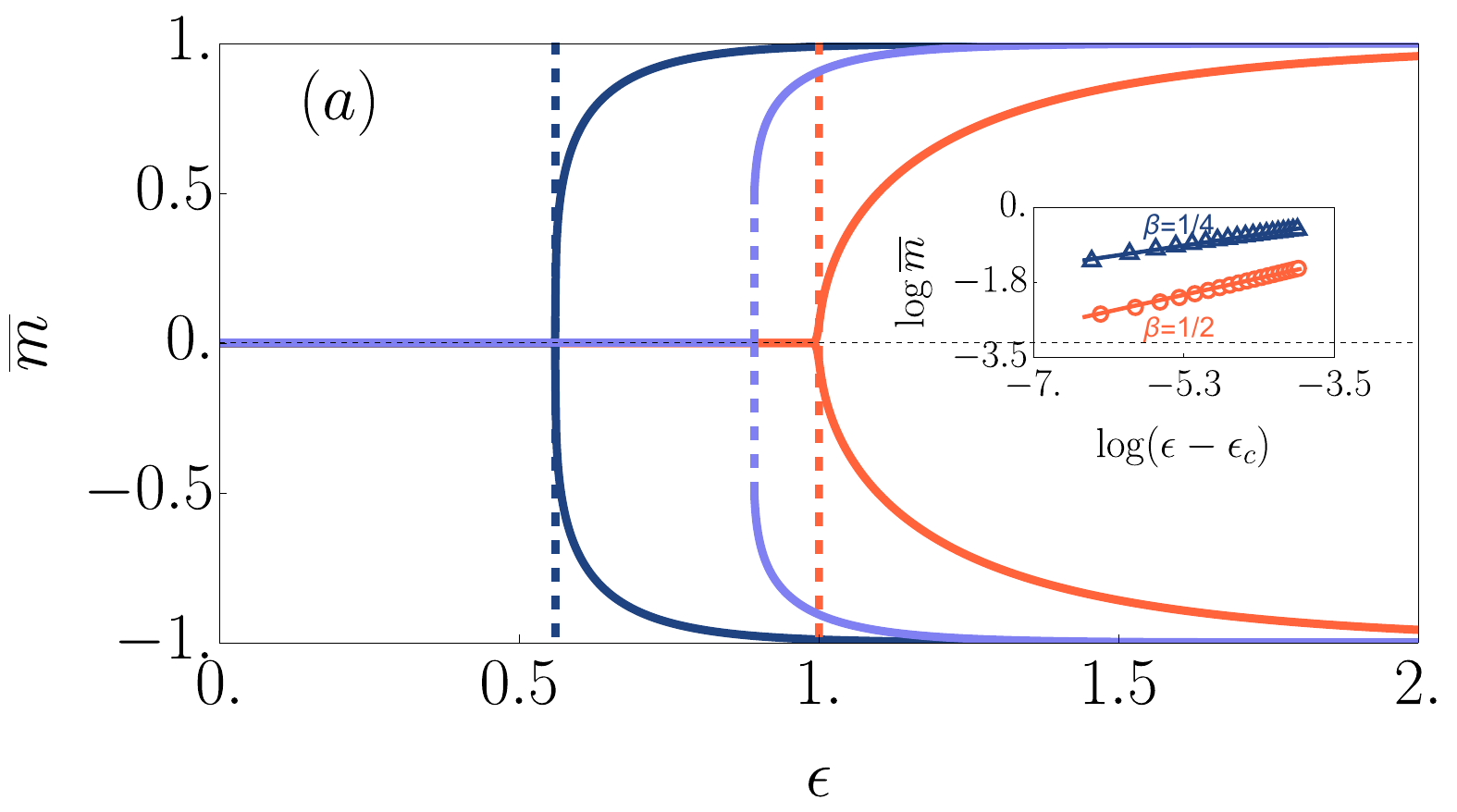}
    \caption{Global magnetization $\overline{m}$ versus $\epsilon$ for the symmetric external parameters ($\theta_1=\theta_2=0$) and different $\lambda_\nu$'s.  Vertical
    dashed lines denote the corresponding transition points. The orange ($\lambda_1=16.2$, $\lambda_2=3$), light blue ($\lambda_1=0.5$, $\lambda_2=-0.69$) and dark blue ($\lambda_1=0.5$, $\lambda_2=1.31$) curves correspond to the second-order, first-order and tricritical transitions, respectively. Inset:
    Log-log plot of $\overline{m}$ versus $\epsilon-\epsilon_c$, whose slopes
    are consistent with  $1/2$ and $1/4$, respectively. Parameters: $\beta_1=6,\beta_2=1,\gamma=1$ and $\delta_1=\delta_2=1$.  }
    \label{mag_Sym}
\end{figure}

\subsection{Symmetric external parameters} The analysis of the symmetric case $\theta_\nu=0$ is more revealing 
because its interplay with other  parameters can give rise to different behaviors, such as   tricritical points. 
From Eqs.~(\ref{sol1_Mag}) and (\ref{sol2_Mag}) as $\kappa\rightarrow \infty$, the steady $\overline{m}$ is given by 
\begin{equation}
    \overline{m}=
    \frac{
         e^{\beta _2 \delta _2 \lambda _2}\,\sinh(\beta _1\epsilon\overline{m})
        +e^{\beta _1 \delta _1 \lambda _1}\,\sinh(\beta _2\epsilon\overline{m})}{e^{\beta _2 \delta _2 \lambda _2}\,\cosh(\beta _1\epsilon\overline{m})
        +e^{\beta _1 \delta _1 \lambda _1}\,\cosh(\beta _2\epsilon\overline{m})
    }. \label{sym}
\end{equation}
It is immediate to see that Eq.~(\ref{sym}) always has the $\overline{m}=0$ as solution,  which is stable for small $\beta_\nu \epsilon$'s. Conversely, the system also admits $\overline{m}\neq 0$ for large  $\beta_\nu \epsilon$'s.  To locate the critical
point $\epsilon_c$ one    expands Eq.~(\ref{sym})  around $\overline{m}=0$, given by
$0=A_1(\epsilon-\epsilon_c)\,\overline{m} + A_3 \overline{m}^3 + A_5 \overline{m}^5 + \cdots ,$ where $\epsilon_c$ reads 
\begin{equation}
    \epsilon_c =
\frac{e^{\beta_1\delta_1\lambda_1}+e^{\beta_2\delta_2\lambda_2}}
         {\beta_1 e^{\beta_2\delta_2\lambda_2}
         +\beta_2 e^{\beta_1\delta_1\lambda_1}},
         \label{crit_Sym}
\end{equation}
and all coefficients of even and odd powers are zero and are listed in Appendix, respectively. 
 As before, the criticality occur when both $A_3<0$ and $A_5<0$, leading to the usual mean-field critical behavior
 $\overline{m}\sim(\epsilon-\epsilon_c)^{\beta_c}$,
 where $\beta_c=1/2$, akin to the antisymmetric case. Conversely,
the system undergoes a discontinuous phase transition when  $A_3>0$ in which
a tricritical point $\epsilon_t$, obtained as $A_3=0$,  marks
the change from both cases. In such case, the magnetization behaves as
$\overline{m}\sim(\epsilon-\epsilon_t)^{\beta_t}$,
where $\beta_t=1/4$ and the condition for the tricriticality is listed
below
    \begin{equation}
    \beta_2\delta_2\lambda_2-\beta_1\delta_1\lambda_1=\log \left[\frac{\beta_2}{2\beta_1}{\left(\frac{\beta_1}{\beta_2}-3\right)}+f(\beta_1,\beta_2)\right],
    \label{tric_condition}
    \end{equation}    
with 
\begin{equation}
    f(\beta_1,\beta_2)=\frac{1}{4}\left(\frac{\beta_2}{\beta_1}-1\right)^3\left[1-\sqrt{1-\frac{12\beta_1^2}{\beta_2^2\left(1-\frac{\beta_1}{\beta_2}\right)^4}}\right],
\end{equation}
where the condition $\beta_1/\beta_2>5.27$ ensures real values.
Fig.~\ref{mag_Sym} illustrates continuous, discontinuous and the tricritical
point for some specific parameter values. Although the emergence of tricritical
point has been reported in the literature of equilibrium systems \cite{salinas2001introduction,yeomans1992statistical}, it commonly requires the inclusion of second and
third nearest neighbor interactions \cite{dosSantosFilho,10.1103/physrevb.84.134422} or at least three kinds of spins per site \cite{beg}.
Our present study unveils that the inclusion of genuine nonequilibrium ingredients
(different temperatures and a symmetric fields) gives rise to similar phenomena
for the minimum  Ising model with nearest neighbor interactions.

\begin{figure}[htb!]
    \centering
    \includegraphics[width=.95\linewidth]{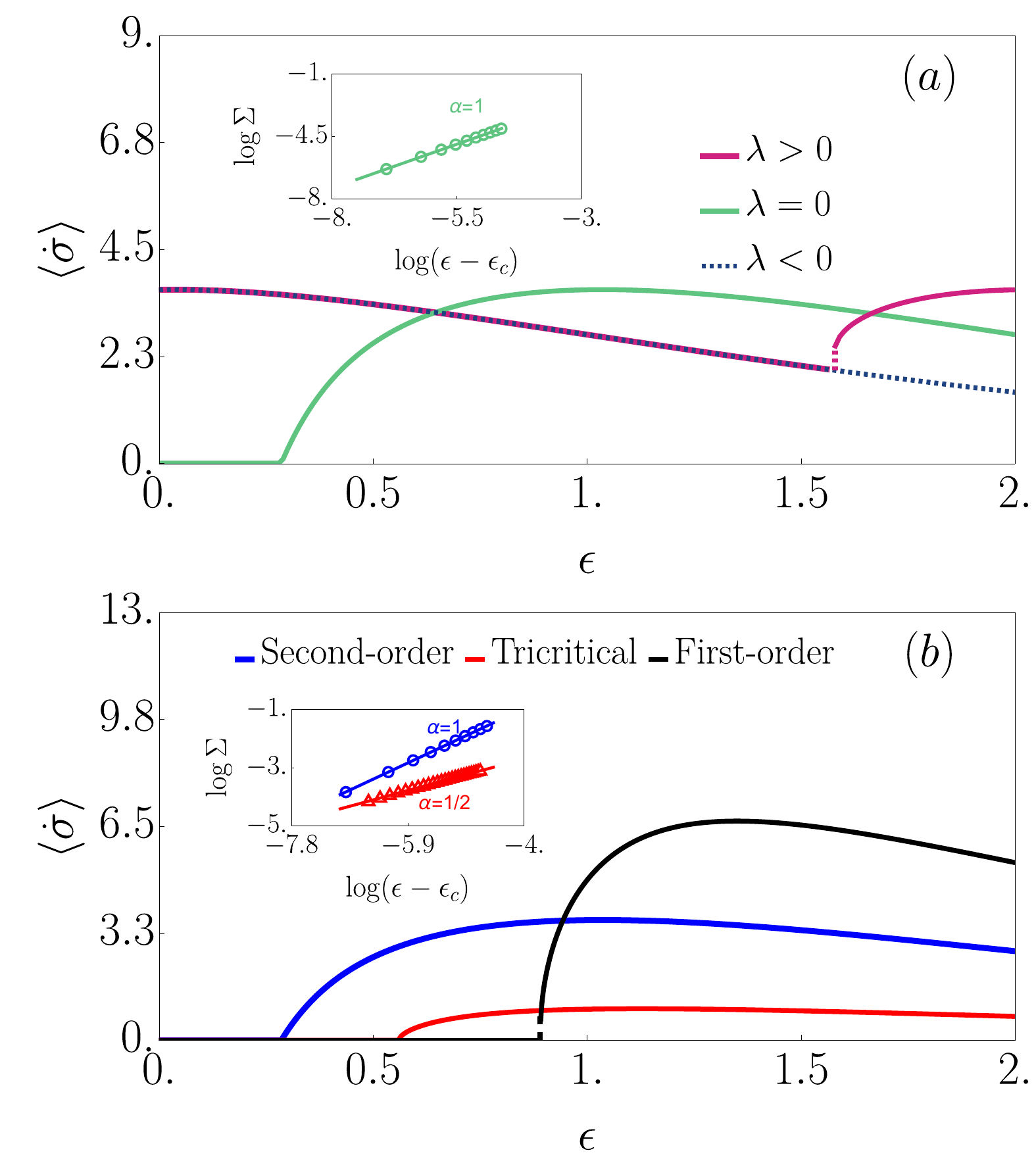}
    \caption{The entropy production $\mom{\dot{\sigma}}$ versus $\epsilon$ for the antisymmetric (a) ($\delta_\nu=0,\theta_\nu=1$\text{ and }$\lambda=0,\pm1$) and symmetric (b) ($\delta_\nu=1,\theta_\nu=0$) parameters in the fast switching limit. In
 $(b)$, we set  $(\lambda_1,\lambda_2)$ as $(0,0)$, $(0.5,-0.69)$ and $(0.5,1.31)$ for the second-order, first-order and tricritical transitions, respectively. Parameters: $\beta_1=6,\beta_2=1,\gamma=1$.}
    \label{SS_Simul}
\end{figure}

We also investigate the behavior of entropy production
at different phase transition regimes. The disordered phase, yielding
for $\epsilon<\epsilon_c$ is marked by different values of $ \mom{\dot{\sigma}}=\mom{\dot{\sigma}}_0$. 
 While $\mom{\dot{\sigma}}_0=0$ for the symmetric case,  it is different for antisymmetric ones and given by 
\begin{equation}
    \mom{\dot{\sigma}}_{0}=2\gamma(\beta_1\theta_1\lambda_1-\beta_2\theta_2\lambda_2)\frac{\sinh\left[\frac{(\beta_1\theta_1\lambda_1-\beta_2\theta_2\lambda_2)}{2}\right]}{\cosh\left[\frac{(\beta_1\theta_1\lambda_1+\beta_2\theta_2\lambda_2)}{2}\right]}.
    \label{epd}
\end{equation}
By combining above expression with Eq.~(\ref{cond_Anti}), one obtains the simpler expression for $\mom{\dot{\sigma}}_{0}=4 \gamma\beta _\nu\theta_\nu \lambda _\nu \sinh \left(\beta _\nu \theta _\nu
   \lambda _\nu\right)$ for $\nu=1\text{ or }2$.
   The entropy production also behaves differently
   in both cases at the criticality. At the vicinity of
   the critical point, one expects a behavior of type \cite{tome2006,mamede2025exactmappingnonequilibriumequilibrium} 
   \begin{equation}
       \Sigma\equiv \mom{\dot{\sigma}}-\mom{\dot{\sigma}}_{0}\sim(\epsilon-\epsilon_c)^\alpha,
   \end{equation}
with $\alpha$ denoting its critical exponent. In order to obtain $\alpha$
in both cases, we express the entropy production in terms of the order-parameter given by $\mom{\dot{\sigma}}=\mom{\dot{\sigma}}_{0}+c_\sigma {\overline m}^2+...$, where the expression for $c_\sigma$ is given by
\begin{equation}
c_\sigma=\gamma\left(\frac{\beta_1-\beta_2}{\beta_1+\beta_2}\right)^2
\left[
 \cosh(\Phi_\nu)
-\frac{2\beta_1\beta_2 \Phi_\nu }{(\beta_1-\beta_2)^2}\sinh(\Phi_\nu)
\right],
\end{equation}
for the antisymmetric case, where $\Phi_\nu\equiv \beta_\nu\theta_\nu\lambda_\nu$ and $\nu=1\text{ or }2$, and
\begin{equation}
    c_\sigma=\frac{\gamma
    \left(\beta _1-\beta _2\right){}^2}{2}\left[\frac{e^{\beta _1 \delta _1 \lambda _1}+e^{\beta _2 \delta _2 \lambda _2}}{ \left(\beta _1 e^{\beta _2 \delta _2
   \lambda _2}+\beta _2 e^{\beta _1 \delta _1 \lambda _1}\right){}^2}\right],
\end{equation}
for the symmetric case.
From the order-parameter behaviors, we obtains $\alpha=2\beta_c=1$
at the criticality and $\alpha=2\beta_t=1/2$ at the tricriticality.
The order parameter jump also manifests in the behavior of the  entropy production, jumping from $\mom{\dot{\sigma}}_{0}$
to $\mom{\dot{\sigma}}$ at discontinuous phase transitions.
Fig. \ref{SS_Simul} depicts aforementioned different  behaviors of the entropy production for both symmetric (bottom) and antisymmetric (top) cases, whose critical behavior is consistent with values as before.

%and plateaus
%for $\epsilon<\epsilon_c$ at top panels follow Eq.~(\ref{epd}). 

%\IM{Symmetric drivings modify transition rates without affecting their ratio, and therefore cannot be written as differences of a state function. As a result, they cannot be absorbed into an effective Hamiltonian as in the case of Eq.\eqref{steady_state_Anti}, leading to persistent probability currents and preventing a Boltzmann–Gibbs description.}
 
\section{Non simultaneous contact between baths}\label{Finite_Hoop}
We  now advance beyond the simultaneous case 
by considering finite $\kappa$. 
Although transition points can be exactly obtained
via the analysis of steady state solutions of Eqs.~\eqref{sol1_Mag}-\eqref{sol2_Mag}, there is
no closed form for them, except to the case $\lambda_1=\lambda_2=0$. In such case, the criticality is given by the following expression \begin{equation}
(\beta_1+\beta_2)\epsilon_c=
\frac{4{(\kappa+1)}}
{\kappa+2
+
\sqrt{\kappa^2
+4\left(\frac{\beta_1-\beta_2}{\beta_1+\beta_2}\right)^2(\kappa+1)}
}. 
\label{ec_nodriving_kappa}
\end{equation} It is immediate to see that Eq.~(\ref{ec_nodriving_kappa}) reduces to Eq.~(\ref{cond_Anti}) as $\kappa\rightarrow \infty$ and deviates meaningfully as $\kappa$ is finite, where $(\beta_1+\beta_2)\epsilon_c \approx 2\left[1-\kappa^{-1}(\beta_1-\beta_2)^2(\beta_1+\beta_2)^{-2}) \right]$ as $\kappa\gg1$.

\begin{figure}[htb!]
    \centering\textbf{}
    \includegraphics[width=1\linewidth]{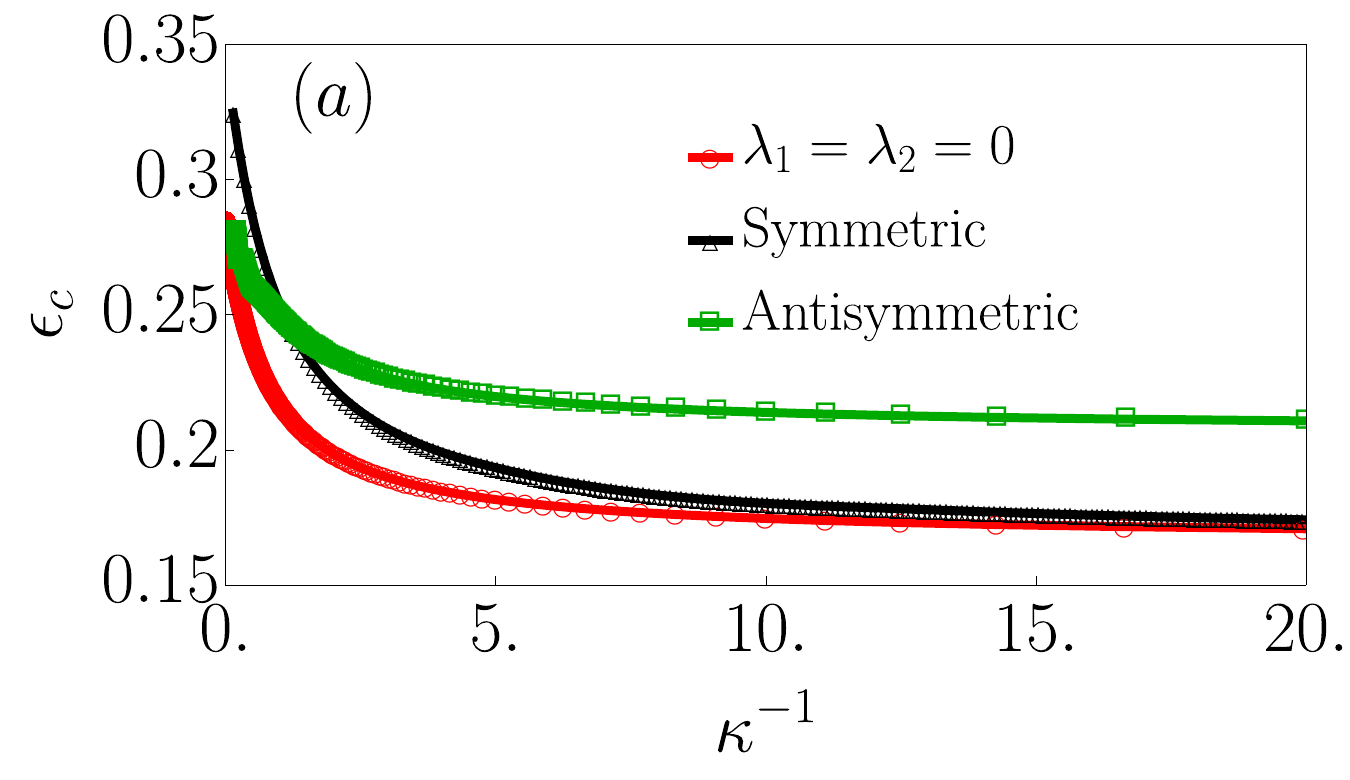}
    \includegraphics[width=1\linewidth]{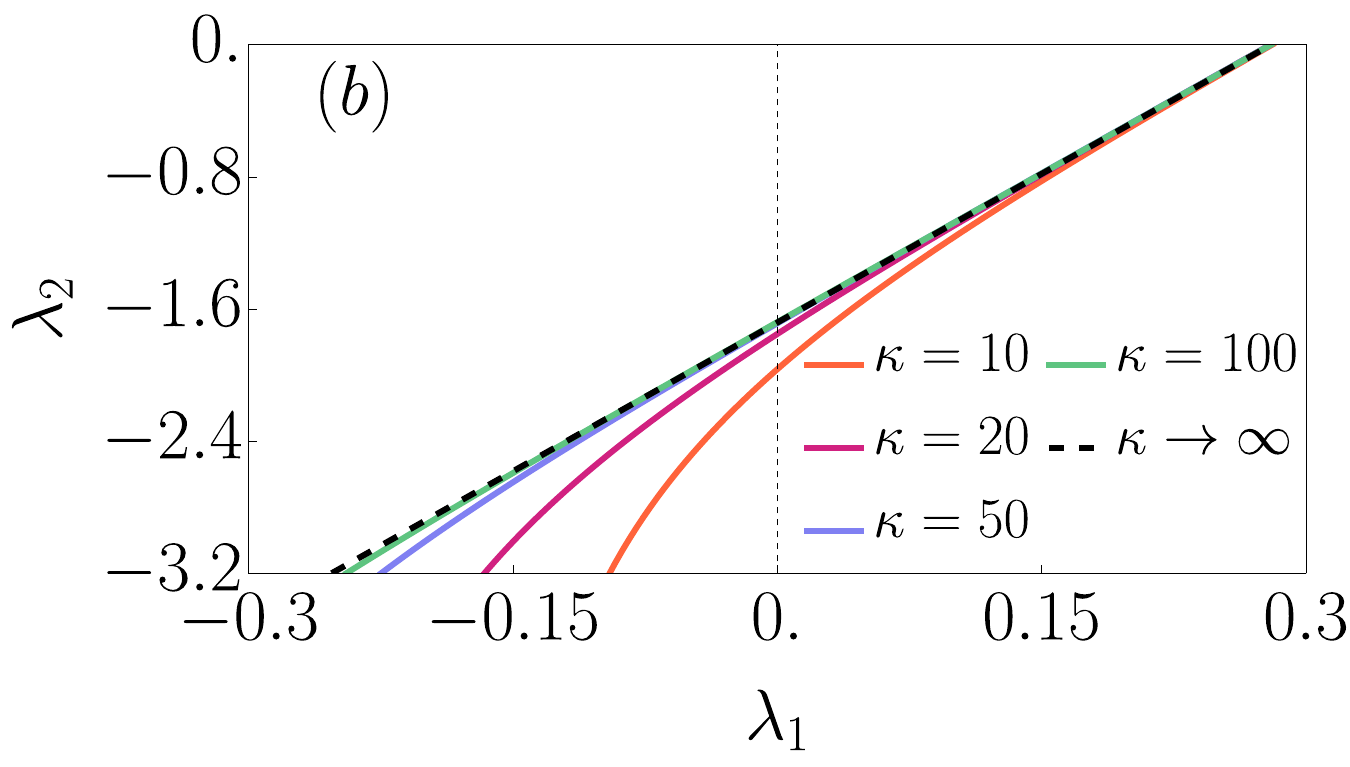}
    \includegraphics[width=1\linewidth]{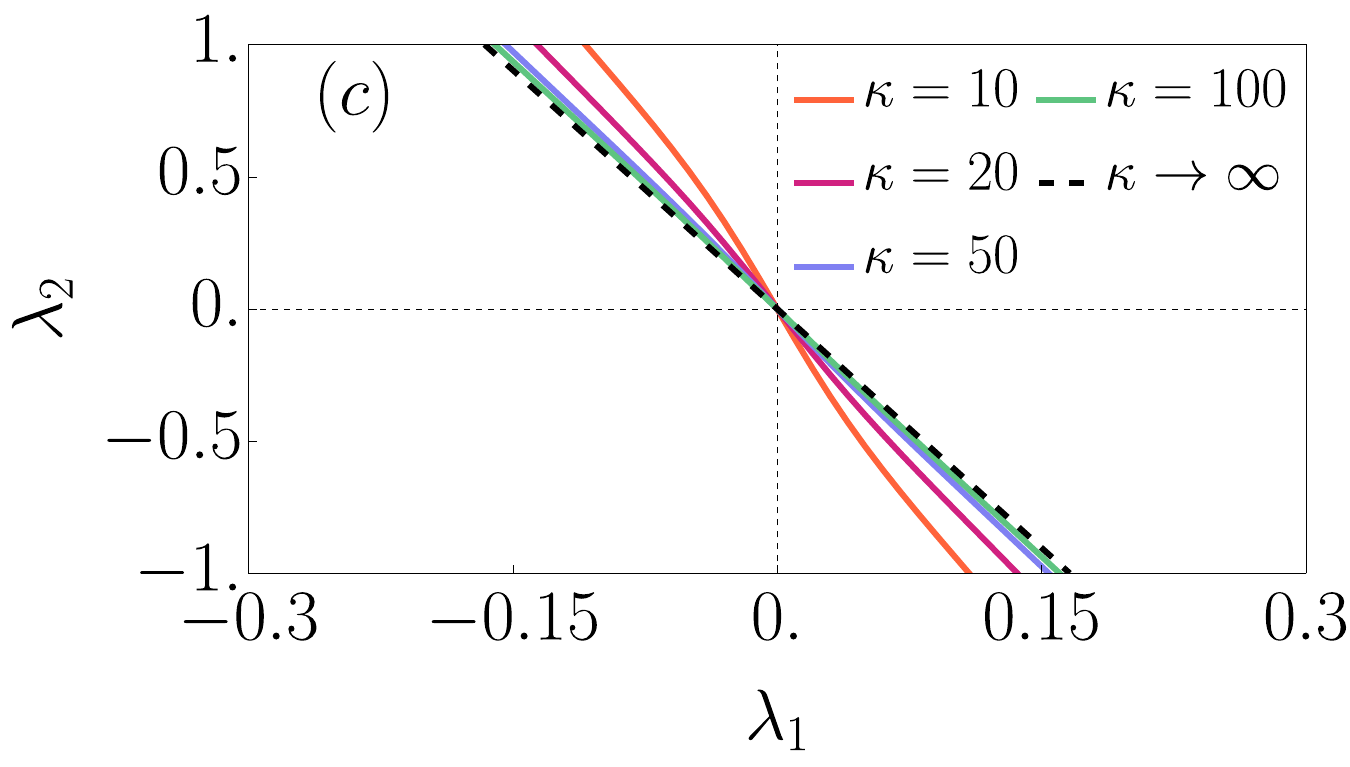}
    \caption{Depiction of phase transitions for  finite switchings between thermal baths.  $(a)$ shows transition points $\epsilon_c$ versus $\kappa^{-1}$  for the absence ($\circ$)(from Eq.\eqref{ec_nodriving_kappa}),
    symmetric (triangles), respectively. Panels $(b)$ and $(c)$ show  tricritical   and critical lines $\lambda_1\times\lambda_2$ for different $\kappa$'s,
    respectively, for $\delta_1=\delta_2=1$ and $\lambda_1=\lambda_2=0.1$ (symmetric) and $\theta_1=\theta_2=1$ and $\lambda_1=-\beta_2\lambda_2/\beta_1$ with $\lambda_2=0.1$ (antisymmetric). They approach to Eqs.~(\ref{tric_condition}) and  \eqref{cond_Anti} as $\kappa\rightarrow\infty$. Parameters: $\gamma=1$, $\beta_1=6$, $\beta_2=1$.}
    \label{Hopping_Fig}
\end{figure}

In all cases, results are qualitatively similar to the simultaneous case, above all the set of  critical exponents and the fact they approach  as $\kappa$ is large. However, there are some remarkable differences. The former is the  fact that phase transitions are always
discontinuous when parameters are chosen according to Eq.~(\ref{cond_Anti}) 
and $\kappa$ is finite, 
being critical only as $\kappa\rightarrow \infty$. 
Unlike the simultaneous case, 
 tricritical lines  deviate
 of the linear form given by Eq.~\eqref{tric_condition} as  $\kappa$
 is finite,  approaching it as $\kappa\gg1$.
Fig.~\ref{Hopping_Fig} exemplifies such findings for finite $\kappa$
but parameters chosen according to Eq.~(\ref{cond_Anti}) and (\ref{tric_condition}). 

%The same behavior
% holds for the  criticality condition for the antisymmetric case in panel %$(c)$. Although they
% are different and present different classifications, being discontinuous as %$\kappa$ is finite from  Eq.~(\ref{cond_Anti}), they approach as %$\kappa\gg1$ and become critical. 

%For the antisymmetric drivings, the corresponding lines in the $\lambda_1 \times \lambda_2$ parameter space were obtained numerically and are displayed in panel $(c)$ of Fig.~\ref{Hopping_Fig}, where for any $\kappa\lessapprox 10^4$ we got first-order phase transitions. Once again, a pronounced dependence on the switching term $\kappa$ is observed, with the numerical curves approaching the condition expressed in Eq.~\eqref{cond_Anti} as $\kappa$ increases.

\section{Conclusions}
The generic properties  of the nonequilibrium Ising model
in contact with two thermal reservoirs has been analyzied. Our study has taken into account  simultaneous and non-simultaneous contact between thermal reservoirs for generic  symmetric and antisymmetric external parameters.
Different findings have been uncovered. 
%\sout{For the fast switching between thermal reservoirs, the probability distribution is Boltzmann-Gibbs like, irrespectively the model parameters (temperatures, external parameters and interaction strength) for antisymmetric external parameters and deviating
%from it for symmetric external parameters and simultaneous contact between thermal baths.} 
For  fast switchings between thermal reservoirs, the probability distribution approaches to Boltzmann–Gibbs form, independently of the model parameters (temperatures, external parameters, and interaction strength) for antisymmetric external parameters. In contrast, symmetric external parameters always deviate from this form. While critical and discontinuous phase transitions exist in the former case, a tricritical behavior is presented for latter case. Expressions for transition points
are remarkably different in both cases, signed by a bilinear relation 
for antisymmetric parameters and deviating from this form for symmetric
parameters.
All  different phase transitions and classifications are also presented results for non-simultaneous contact between thermal reservoirs, revealing not only the robustness of such phenomena but also the reliability
of simultaneous contact between thermal baths.
In summary, the inclusion of different nonequilibrium ingredients
lead to a variety of phase transitions and critical behaviors.
As potential perspectives,  we cite the investigation of other complex many-body settings (e.g., systems with energetic frustration) as well as the role of asymmetric stochastic and deterministic switching between the hot and cold reservoirs. Allowing different switching rates between baths will break time-reversal symmetry more strongly and may reveal richer phase structure or additional ways to control dissipation and fluctuations.

%\appendix

\setcounter{equation}{0}
\renewcommand{\theequation}{A\arabic{equation}}
\makeatletter
\renewcommand{\theHequation}{A\arabic{equation}}
\makeatother

\section{Appendix: Entropy production derivation}\label{entropy_prod_section}
At the non-equilibrium steady state, the entropy production in the limit of $N\to\infty$ obeys the following Schnakenberg relation~\cite{PhysRevLett.108.020601}

\begin{equation}
    \mom{\dot{{\sigma}}}=\sum_{\nu}[\omega^{(\nu)}_{+-}\overline{x}^{(\nu)}_{-}-\omega^{(\nu)}_{-+}\overline{x}^{(\nu)}_{+}]\log\left(\frac{\omega^{(\nu)}_{+-}}{\omega^{(\nu)}_{-+}}\right),
\end{equation}
with $\overline{x}^{(\nu)}_{i}$ is the density of states at the steady state and where, given the symmetric switching between baths, only the spin-exchange transition rates are non-zero. By using the the transition rates definitions from Eqs.\eqref{Omega_pm}-\eqref{Omega_mp}, along with the definitions at the steady state
\begin{equation}
    \overline{x}^{(\nu)}_{+}=\frac{1+\overline{m}_\nu}{2},\qquad\overline{x}^{(\nu)}_{-}=\frac{1-\overline{m}_\nu}{2},
\end{equation}
the term $\log(\omega^{(\nu)}_{+-}/\omega^{(\nu)}_{-+})$, 
is straightforwardly reduced to $2\overline{H}_\nu$. Hence, we shall obtain the following 
\begin{equation}
    \mom{\dot{{\sigma}}}=\sum_{\nu}\overline{H}_\nu(\omega^{(\nu)}_{+-}-\omega^{(\nu)}_{-+})\left[1+\overline{m}_\nu\left(\frac{\omega^{(\nu)}_{+-}+\omega^{(\nu)}_{-+}}{\omega^{(\nu)}_{+-}-\omega^{(\nu)}_{-+}}\right)\right].
\end{equation}
Finally, by using again the definitions from Eqs.\eqref{Omega_pm}-\eqref{Omega_mp}, we shall get, under simplifications, that
\begin{equation}
    \frac{\omega^{(\nu)}_{+-}+\omega^{(\nu)}_{-+}}{\omega^{(\nu)}_{+-}-\omega^{(\nu)}_{-+}}=-\coth\overline{H}_\nu,
\end{equation}
and, moreover, obtaining
\begin{equation}
    \mom{\dot{{\sigma}}}=\sum_{\nu}\overline{H}_\nu(\omega^{(\nu)}_{+-}-\omega^{(\nu)}_{-+})\left(1-\overline{m}_\nu\coth\overline{H}_\nu\right),
\end{equation}
being the exactly Eq.\eqref{EP_macro} in the main text.

\section{Appendix: Coefficients of the order-parameter expansion for antisymmetric and symmetric external parameters}
As shown in the main text,  the critical behaviors can be obtained by expanding order-parameter expressions in power series in such a way that $0=A_0+A_1(\epsilon-\epsilon_c)\overline{m}+A_2\overline{m}^2+A_3\overline{m}^3+A_4\overline{m}^4+A_5\overline{m}^5+\dots$, where  coefficients  $A_i$'s are listed below
\begin{align}
A_0 &= -4\gamma \sinh\left(\dfrac{\beta_1\theta_1\lambda_1 + \beta_2\theta_2\lambda_2}{2}\right) \cosh\left(\dfrac{\beta_1\theta_1\lambda_1- \beta_2\theta_2\lambda_2}{2}\right),
\nonumber\\
A_1 &= 2\gamma \sum_{i=1}^{2} \beta_i \cosh(\beta_i\theta_i\lambda_i),
\\
A_2 &= 8\gamma \sinh\left(\dfrac{\beta_1\theta_1\lambda_1 + \beta_2\theta_2\lambda_2}{2}\right) \cosh\left(\dfrac{\beta_1\theta_1\lambda_1- \beta_2\theta_2\lambda_2}{2}\right),
\nonumber\\
A_3 &= -\frac{4\gamma}{3(\beta_1+\beta_2)^3}
\sum_{i=1}^{2}
\beta_i^2(\beta_i+3\beta_j)
\cosh(\beta_i\theta_i\lambda_i),
\nonumber
\end{align}

%The directly format of all the coefficients strongly depends on the nature of the external parameter. In other words, as described in the main text, we can split this limit situation in sub-cases of antisymmetrical ($\alpha^{(\nu)}_{-+}=-\alpha^{(\nu)}_{+-}$) or symmetrical ($\alpha^{(\nu)}_{-+}=\alpha^{(\nu)}_{+-}$) external parameters. 
%For the first, due to the $\mathbb
%Z_2$ symmetry, in the order-parameter expansion we shall obtain conditions for the external terms in order to not break the criticality. Performing the expansion, we get that
\begin{comment}
\begin{align}
    A_0&=-4 \gamma  \sinh \left(\frac {\beta _1 \theta _1 \lambda
   _1+\beta _2 \theta _2 \lambda _2}{2}\right) \cosh
   \left(\frac {\beta _1 \theta _1 \lambda
   _1-\beta _2 \theta _2 \lambda _2}{2}\right),\nonumber\\
   A_1&=2 \gamma  \left[\beta _1 \cosh \left(\beta _1 \theta _1 \lambda
   _1\right)+\beta _2 \cosh \left(\beta _2 \theta _2 \lambda
   _2\right)\right],\\
   A_2&=\frac{8\gamma\beta_1\beta_2}{\beta_1\beta_2}\sinh \left(\frac {\beta _1 \theta _1 \lambda
   _1+\beta _2 \theta _2 \lambda _2}{2}\right) \cosh
   \left(\frac {\beta _1 \theta _1 \lambda
   _1-\beta _2 \theta _2 \lambda _2}{2}\right),\nonumber\\
   A_3&=\frac{4 \gamma  \left(\left(\beta _1+3 \beta _2\right) \beta _1^2 \left(-\cosh \left(\beta _1 \theta _1 \lambda _1\right)\right)-\beta _2^2 \left(3 \beta _1+\beta
   _2\right) \cosh \left(\beta _2 \theta _2 \lambda _2\right)\right)}{3 \left(\beta _1+\beta _2\right){}^3}
\end{align}
\end{comment}
and so on, for $j\neq i$ and the antisymmetric case, whereas
\begin{comment}
\begin{align}
    A_1&=2 \gamma  \left(\beta _1 e^{-\beta _1 \delta _1 \lambda _1}+\beta _2
   e^{-\beta _2 \delta _2 \lambda _2}\right),\\
   A_3&=\frac{1}{3} \gamma  \epsilon_c ^2 \left[\beta _1^2 \left(\beta _1\epsilon_c
   -3\right) e^{-\beta _1 \delta _1 \lambda _1}+\beta _2^2 \left(\beta
   _2\epsilon_c -3\right) e^{-\beta _2 \delta _2 \lambda _2}\right],\nonumber\\
   A_5&=\frac{1}{60} \gamma \epsilon_c ^4 \left(\beta _1^4 \left(\beta _1
  \epsilon_c -5\right) e^{-\beta _1 \delta _1 \lambda _1}+\beta _2^4
   \left(\beta _2\epsilon_c -5\right) e^{-\beta _2 \delta _2 \lambda
   _2}\right),\nonumber\\
\end{align}
\end{comment}
\begin{align}
A_1 &= 2\gamma \sum_{i=1}^{2} \beta_i e^{-\beta_i\delta_i\lambda_i},
\nonumber\\
A_3 &= \frac{\gamma \epsilon_c^2}{3}
\sum_{i=1}^{2}
\beta_i^2(\beta_i \epsilon_c - 3)
e^{-\beta_i\delta_i\lambda_i},\\
A_5 &= \frac{\gamma \epsilon_c^4}{60}
\sum_{i=1}^{2}
\beta_i^4(\beta_i \epsilon_c - 5)
e^{-\beta_i\delta_i\lambda_i}.
\nonumber
\end{align}
for the symmetric case,  evaluated at  $\epsilon_c$ given by Eq.~\eqref{crit_Sym} in the main text.
\acknowledgments
We acknowledge the financial support from Brazilian agencies CNPq and FAPESP under grants 2023/17704-2, 2024/08157-0, 2024/03763-0, 2022/15453-0. This study was supported by the Special Research Fund
(BOF) of Hasselt University under Grant No. BOF25BL12.

\bibliographystyle{eplbib}
\bibliography{refs_bib}
\end{document}